\documentclass[epj]{svjour}
\usepackage{amsmath,amssymb}
\usepackage{bm}
\usepackage{graphicx}
\usepackage{epsfig}
\usepackage{psfrag}

\newcommand{\bd}{\bm}

\begin{document}

\title{Microscopic spin-wave theory for yttrium-iron garnet films}

\author{Andreas Kreisel \and Francesca Sauli \and Lorenz Bartosch \and Peter Kopietz}

\institute{Institut f\"{u}r Theoretische Physik, Universit\"{a}t
  Frankfurt,  Max-von-Laue Strasse 1, 60438 Frankfurt, Germany}
\date{August 25, 2009}

 \abstract{
Motivated by recent experiments on thin films of the ferromagnetic insulator 
yttrium-iron garnet (YIG), we have developed an efficient 
microscopic approach to calculate
the spin-wave spectra of these systems.
We model the experimentally relevant magnon band of YIG using an
effective quantum Heisenberg model on a cubic lattice
with ferromagnetic nearest neighbour exchange and long-range dipole-dipole interactions.
After a bosonization of the spin degrees of freedom 
via a Holstein-Primakoff transformation 
and a truncation at quadratic order in the bosons, we obtain the
spin-wave spectra for experimentally relevant parameters
without further approximation by
numerical diagonalization, using efficient Ewald summation techniques to carry
out the dipolar sums.
We compare our numerical results with two different analytic approximations and with
predictions based on the phenomenological Landau-Lifshitz equation.}

\maketitle
\section{Introduction}

In a recent series of experiments~\cite{Demokritov06,Demidov07,Dzyapko07,Demidov08,Demokritov08} Demokritov and co-workers 
discovered strong correlations of highly occupied magnon states 
in thin  films of the magnetic insulator
yttrium-iron garnet (YIG) with stoichiometric formula Y$_3$Fe$_2$(FeO$_4$)$_3$.
They suggested an interpretation of their results
in terms of Bose-Einstein condensation of magnons at room temperature.
For a proper interpretation of these experiments,
a peculiar feature of the energy dispersion $E_{\bd{k}}$ of the relevant
magnon band in  finite YIG films is important:
due to a subtle interplay between finite-size effects, short-range
exchange interactions, and long-range dipole-dipole interactions,
$E_{\bd{k}}$ exhibits a local minimum at a finite wave-vector $\bd{k}_{\text{min}}$,
for a certain range of orientations of the
external magnetic field $\bd{H}_e$ relative to the sample.
The existence of such a dispersion minimum has been predicted
by Kalinikos and Slavin~\cite{Kalinikos86,Cottam94}  within a phenomenological approach
based  on the Landau-Lifshitz equation.
Unfortunately, such a phenomenological approach does not provide a
microscopic understanding of correlation effects,
which might be  important to explain some aspects
of experiments probing the non-equilibrium behaviour of the
magnon gas in YIG~\cite{Demokritov06,Demidov07,Dzyapko07,Demidov08,Demokritov08,Serga07,Schaefer08,Chumak09,Neumann08}.
This has motivated us to study this problem within the framework
of the usual $1/S$-expansion for ordered quantum spin systems, 
which is based on the bosonization of an effective microscopic Heisenberg model
using either the Holstein-Primakoff~\cite{Holstein40} or the 
Dyson-Maleev transformation~\cite{Dyson56,Maleev57},
and the subsequent classification of the interaction processes  
in powers of the small parameter $1/S$.

The $1/S$-expansion
has been extremely successful to understand spin-wave interactions
in ordered magnets~\cite{Akhiezer68,Mattis06}.
Previously, several authors have used this approach to calculate spin-wave spectra
in ultrathin ferromagnetic films with exchange and 
dipole-dipole interactions~\cite{Erickson91a,Erickson91,Pereira99,Rezende09}. 
Moreover, interaction effects such as energy shifts and damping of
spin-waves in thin films have also been calculated within the 
$1/S$-expansion~\cite{Costa98,Costa00}.
However, in order to apply this approach to realistic models for
experimentally relevant YIG films with a thickness of
a few microns (corresponding to
a few thousand lattice spacings), 
one has to evaluate numerically rather large dipolar sums~\cite{Benson69} 
to set up the secular matrix
whose eigenvalues determine the magnon modes, see
Eq.~(\ref{eq:det}) below.
We use here an efficient Ewald summation technique~\cite{Ziman79} to carry out these 
summations, which enables us to calculate the spin-wave dispersions of
realistic YIG films.  Given our numerical results, we can assess the
validity of various analytical approximations such as the
uniform mode approximation~\cite{Kalinikos86,Rezende09,Kostylev07} and the lowest eigenmode approximation.

The rest of this paper is organized as follows: 
In Sec.~\ref{sec:Hamiltonian} we 
introduce the effective Heisenberg model which we shall use to describe the
experimentally relevant magnon band in YIG. We set up the $1/S$-expansion and
derive the secular equation which determines the magnon dispersion.
In Sec.~\ref{sec:spectra} we present our results for the magnon spectra of YIG.
We first discuss
our numerical results, which are obtained by evaluating the
roots of the secular determinant without further approximation, using the
Ewald summation technique described in the appendix to 
evaluate the necessary dipolar sums.
We then discuss in Secs.~\ref{subsec:uniform} and \ref{subsec:eigen}
two approximate analytical methods for obtaining the dispersion
of the lowest magnon band. A comparison with our 
numerical results allows us to estimate the accuracy of these approximations.
Finally, in Sec.~\ref{sec:conclusions}
we present our conclusions and give an outlook for further research.

\section{Effective Hamiltonian for YIG films}
\label{sec:Hamiltonian}

Experimental and theoretical research on YIG has a long 
history, as reviewed, for example,  in 
Ref.~\cite{Cherepanov93}.
Some distinct advantages of YIG are that this material can be grown in very pure crystals
and  has a very narrow ferromagnetic resonance line, indicating
very low  spin-wave damping.
Actually,  YIG is a ferrimagnet at accessible magnetic fields and
has a rather complicated crystal structure 
with  space group $Ia3d$ (see
Refs.~\cite{Cherepanov93,Gilleo58}) 
and 20 magnetic ions 
in the primitive cell. 
Fortunately, on the energy scales 
relevant to experiments~\cite{Demokritov06,Demidov07,Dzyapko07,Demidov08,Demokritov08,Serga07,Schaefer08,Chumak09,Neumann08}
only the lowest magnon band is important, so that
we can describe the physical properties of YIG
at room temperature   
in terms of an effective spin $S$ quantum Heisenberg ferromagnet 
on a cubic lattice with lattice spacing \cite{Gilleo58}
 \begin{equation}
 a=12.376\,\text{\AA}\;.
 \end{equation}  
The effective Hamiltonian contains both
exchange and dipole-dipole interactions,
\begin{eqnarray}
 \hat H&=&-\frac 12 \sum_{ij} J_{ij} \bd S_i \cdot \bd S_j -\mu \bd H_e\cdot \sum_i\bd S_i\notag\\
&&-\frac 12 \sum_{ij,i\neq j}\frac{\mu^2}{|\bd R_{ij}|^3} \left[3 (\bd S_i\cdot\hat{\bd R }_{ij})(\bd S_j\cdot\hat{\bd R }_{ij}) -\bd S_i \cdot \bd S_j\right], \notag\\
 \label{eq:hamiltonian}
\end{eqnarray}
where the sums are over the sites $\bd R_i$ 
of the lattice and
$\hat{\bd R}_{ij}=\bd R_{ij}/|\bd R_{ij}|$ are  unit vectors in the direction 
of $\bd R_{ij}=\bd R_i-\bd R_j=x_{ij}\bd e_x+y_{ij}\bd e_y+z_{ij}\bd e_z$. 
Here, $\mu=g\mu_B$ is the magnetic moment associated with the spins, 
where $g$ is the effective $g$-factor and $\mu_B=e\hbar /(2mc)$ is the Bohr magneton.  
The exchange energies $J_{ ij}=J(\bd R_i-\bd R_j)$ decay rapidly with distance, so that
it is sufficient to include only nearest neighbour exchange couplings in 
Eq.~(\ref{eq:hamiltonian}), setting $J_{ ij}   = J$ if
$\bd R_i$ and $\bd R_j$ are nearest neighbours, and 
$J_{ij} =0$ otherwise.
Note that we neglect surface anisotropies which might be present in experiments, especially at the surface of the film which is attached to the substrate.
Experimentally, the material YIG is characterized by its saturation magnetization~\cite{Tittmann73}
 \begin{equation}
 4\pi M_S=1750\,\text{G}, 
 \label{eq:MS}
 \end{equation}
and the exchange stiffness $\rho_{\rm ex}$ of long-wavelength
spin-waves~\cite{Gurevich},
 \begin{equation}
\frac{ \rho_{\rm ex}}{\mu} = \frac{JS a^2}{\mu}\approx 5.17 \times 
 10^{-13} \;\text{Oe\,m}^2.
 \label{eq:rhoex}
 \end{equation}
If we arbitrarily set the effective $g$-factor equal to two~\cite{Tittmann73} so that
$\mu = 2 \mu_B$,
we obtain from Eqs.~(\ref{eq:MS}) and (\ref{eq:rhoex})
for the effective spin 
\begin{equation}
 S=\frac{M_S a^3}{\mu} \approx 14.2,
\end{equation}
and for the nearest neighbour exchange coupling
\begin{equation}
  J = 1.29\,\text{K}.
\end{equation}
Our above estimates for $S$ and $J$ differ slightly
from the values given in
Ref.~\cite{Tupitsyn08}.
Note that the effective spin $S\approx 14.2$ is quite large,
so that an expansion in powers of $1/S$ is justified.
Introducing the dipolar tensor 
$D_{ij}^{\alpha \beta}=D^{\alpha\beta}(\bd R_i-\bd R_j)$,
\begin{eqnarray}
 D_{ij}^{\alpha\beta}&=&(1-\delta_{ij})\frac{\mu^2}{|\bd R_{ij}|^3}\left[3\hat{R}_{ij}^\alpha\hat{R}_{ij}^\beta-\delta^{\alpha\beta}\right]\notag\\
&=&(1-\delta_{ij})\mu^2\frac{\partial^2}{\partial R_{ij}^\alpha\partial 
R_{ij}^\beta} \frac{1}{|\bd R_{ij}|},
\label{eq:defdip}
\end{eqnarray}
we can write our effective Hamiltonian (\ref{eq:hamiltonian}) in the compact form
\begin{equation}
 \hat H=-\frac 12 \sum_{ij}\sum_{\alpha\beta} \left[J_{ij}\delta^{\alpha\beta}+D_{ij}^{\alpha\beta}\right] S_i^\alpha S_j^\beta -h\sum_i S_i^z,
\end{equation}
where the $z$-axis of our coordinate system points into the direction
defined by the magnetic field $\bd{H}_e$ and we have introduced 
the associated Zeeman energy,
 \begin{equation}
 h =\mu |\bd H_e|.
 \end{equation}
We have thus related the set of parameters
$a,S,J, h$ appearing in our effective Hamiltonian to experimentally
measurable quantities.

\begin{figure}[tb]
  \centering
   \includegraphics[width = 70mm]{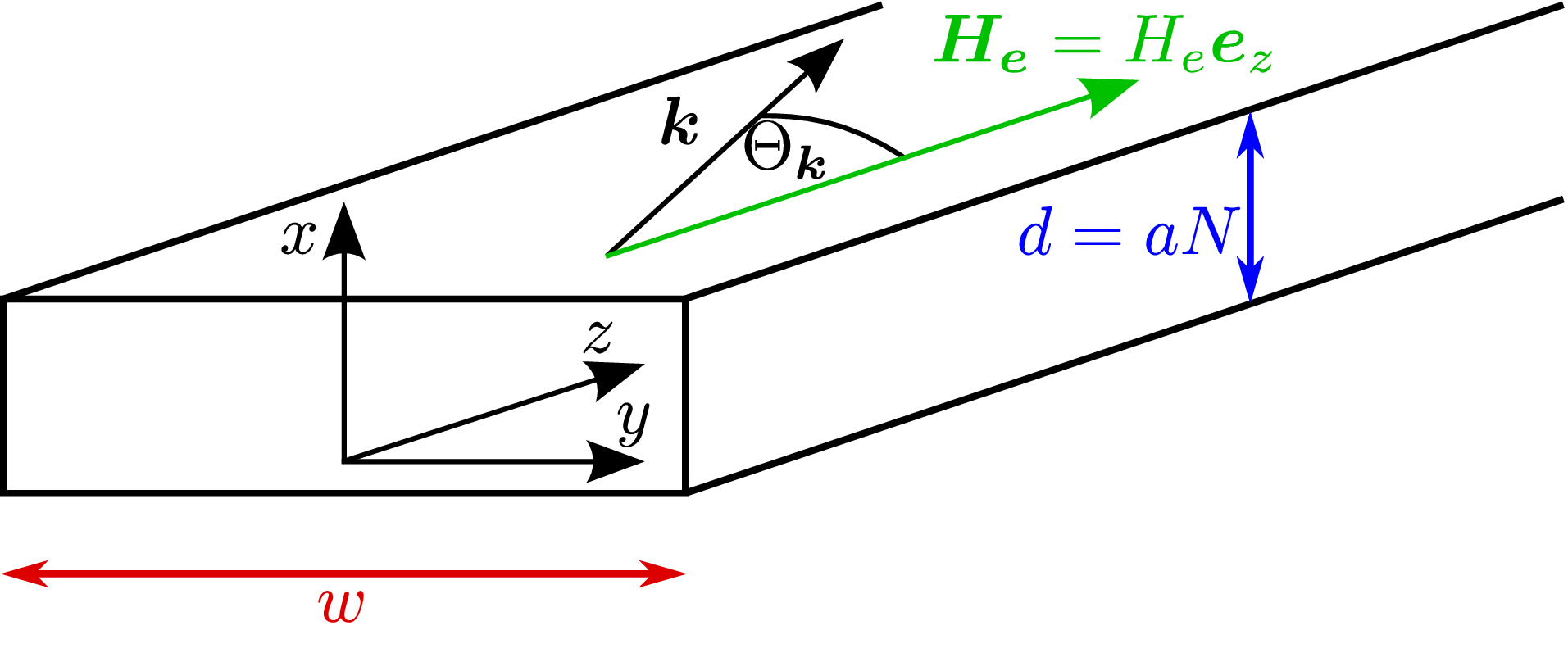}
  \caption{
(Color online)
Orientation of our coordinate system for an infinitely long
stripe of width $w$ and thickness $d$. We assume that the external magnetic field 
$H_e \bd{e}_z$ is parallel
to the long axis (which we call the $z$-axis) of the stripe.
}
\label{fig:stripe}
\end{figure}

To proceed, we restrict ourselves to the description of an 
infinitely long stripe of width $w$ and thickness $d=Na$, consisting of $N$ layers. 
For the stripe geometry shown in Fig. \ref{fig:stripe}  
where the magnetic field points in any direction parallel to the stripe 
the classical groundstate is a saturated ferromagnet. 
Therefore we can expand the Hamiltonian in terms of 
bosonic  operators describing
fluctuations around the classical groundstate, 
using either the Holstein-Primakoff or Dyson-Maleev 
transformation~\cite{Akhiezer68,Mattis06}.
The resulting bosonized spin Hamiltonian is of the form
\begin{equation}
  \hat{H} = H_{0} + \sum_{n=2}^{\infty} \hat{H}_n \; .
 \end{equation}
It turns out that  Holstein-Primakoff and Dyson-Maleev transformations give
different results for $\hat{H}_n$ with $n \geq 4$, but 
the expressions   for  $\hat{H}_2$ and $\hat{H}_3$
are identical in both transformations.
The classical ground state energy is
 \begin{equation}
 H_0 = - \frac{S^2}{2} \sum_{ij} \left[  J_{ij} + D_{ij}^{zz}   + \frac{2 J_{ij} - D_{ij}^{zz}}{2S} \right],
 \end{equation} 
and the quadratic part of the Hamiltonian reads
 \begin{equation}
 \hat{H}_2 = \sum_{ij} \left[ A_{ij} b^{\dagger}_i b_j + \frac{B_{ij}}{2}
 \left( b_i b_j + b^{\dagger}_i b^{\dagger}_j \right)  \right]\; ,
 \label{eq:H2} 
\end{equation}
with
 \begin{subequations}
 \begin{eqnarray}
 A_{ij} & = & \delta_{ij} h + S ( \delta_{ij} \sum_{n} J_{ in} - J_{ ij} )
 \nonumber
 \\ 
 & + & S \left[ \delta_{ij} \sum_{n} D_{in}^{zz}-
\frac{D_{ij}^{xx} + D_{ij}^{yy}}{2} \right],
 \\
 B_{ij} & = & - \frac{S}{2}  \left[ D_{ij}^{xx} - 2 i  D_{ij}^{xy} - D_{ij}^{yy} \right].
 \end{eqnarray}
 \end{subequations}
Since $\hat H_n/S^2 =\mathcal{O}(1/S^{n/2})$ and the effective $S$ is large,
we expect accurate results even if we only keep the first two terms. 
In order to proceed we have to keep in mind that a stripe with thickness $d$ and width $w$ is obviously not translationally invariant, 
so that we cannot simply use a full Fourier transform to diagonalize the Hamiltonian.
Because in the experimentally studied 
samples~\cite{Demokritov06,Demidov07,Dzyapko07,Demidov08,Demokritov08,Serga07,Schaefer08,Chumak09,Neumann08} the width
$w$ of the stripe is much larger than the thickness $d$,
 we can assume that $w$ is practically infinite, 
so that our system can be considered to have discrete translational invariance in the 
$y$- and $z$-directions. We may then partially diagonalize  $\hat{H}_2$ in Eq.~(\ref{eq:H2}) by performing a partial Fourier transformation in
the $yz$-plane. Setting $ \bd R_i = ( x_i , \bd{r}_i )$ with 
$\bd{r}_i = ( y_i , z_i )$, and introducing the two-dimensional 
wave-vector $\bd{k} = ( k_y, k_z )$ in the $yz$-plane,
we expand
 \begin{equation}
 b_i = \frac{1}{\sqrt{N_{y} N_z}} \sum_{ \bd{k}} e^{ i  \bd{k} \cdot \bd{r}_i }
 b_{\bd{k}} ( x_i )\;,
 \end{equation} 
where $N_y$ and $N_z$ is the number of lattice sites in $y$ and $z$
direction. Our Hamiltonian $\hat{H}_2$ in Eq.~(\ref{eq:H2})
takes then the form
 \begin{eqnarray}
 \hat{H}_{2} & = & \sum_{\bd{k}} \sum_{ x_i, x_j} 
 \Bigl [ A_{ \bd{k} } (  x_{ij} )   b^{\dagger}_{ \bd{k} } (  x_i )  b_{ \bd{k} } (  x_j ) 
 \nonumber
 \\ 
& &   +
  \frac{  B_{  \bd{k} } ( x_{ij}) }{2}  b_{ \bd{k} } (x_i )  b_{ -\bd{k} } ( x_j )
  \nonumber
 \\ 
& &  
+   \frac{  B^{ \ast }_{ \bd{k}  } ( x_{ij}) }{2}  b^{\dagger}_{ \bd{k} } ( x_i  ) 
b^{\dagger}_{ - \bd{k} } ( x_j )
 \Bigr],
 \label{eq:H2Gauss}
\end{eqnarray}
with the amplitude factors~\cite{Costa98,Costa00}
\begin{subequations}
 \begin{eqnarray}
  A_{ \bd{k} } (  x_{ij} )  & = & \sum_{ \bd{r} } 
 e^{ - i \bd{k} \cdot \bd{r} } A ( x_i - x_j , \bd{r} )\; ,
 \nonumber\\
&=&SJ_{\bd{k}}(x_{ij})+\delta_{ij}\bigl[ h+S\sum_{n}D_0^{zz}(x_{in})\bigr]\nonumber\\
&&-\frac S2 \bigl[D_{\bd{k}}^{xx}(x_{ij})+D_{\bd{k}}^{yy}(x_{ij})\bigr ] , \\
B_{ \bd{k} } (  x_{ij} )  & = & \sum_{ \bd{r} } 
 e^{ - i \bd{k} \cdot \bd{r} } B ( x_i - x_j , \bd{r} )\nonumber\\
&&\hspace{-1.05cm}=-\frac S2 \bigl[D_{\bd{k}}^{xx}(x_{ij})-2i D_{\bd{k}}^{xy}(x_{ij})-D_{\bd{k}}^{yy}(x_{ij})\bigr],\hspace{1cm}
 \end{eqnarray}
\label{eq:ab}
\end{subequations}
and the exchange matrix
\begin{eqnarray}
 \hspace{-.5cm}J_{\bd{k}}(x_{ij})&=&J \bigl[\delta_{ij}\bigl\{6-\delta_{j1}-\delta_{jN}\notag\\
&& \hspace{-.5cm}-2(\cos (k_ya)+\cos(k_za))\bigr\}-\delta_{i j+1}-\delta_{i j-1}\bigr].
\label{eq:exmatrix}
\end{eqnarray}

\section{Spin-wave spectra of YIG}
\label{sec:spectra}

\subsection{Numerical approach}
\label{subsec:num}

\begin{figure*}[t]
\sidecaption
   \includegraphics[width = 130mm]{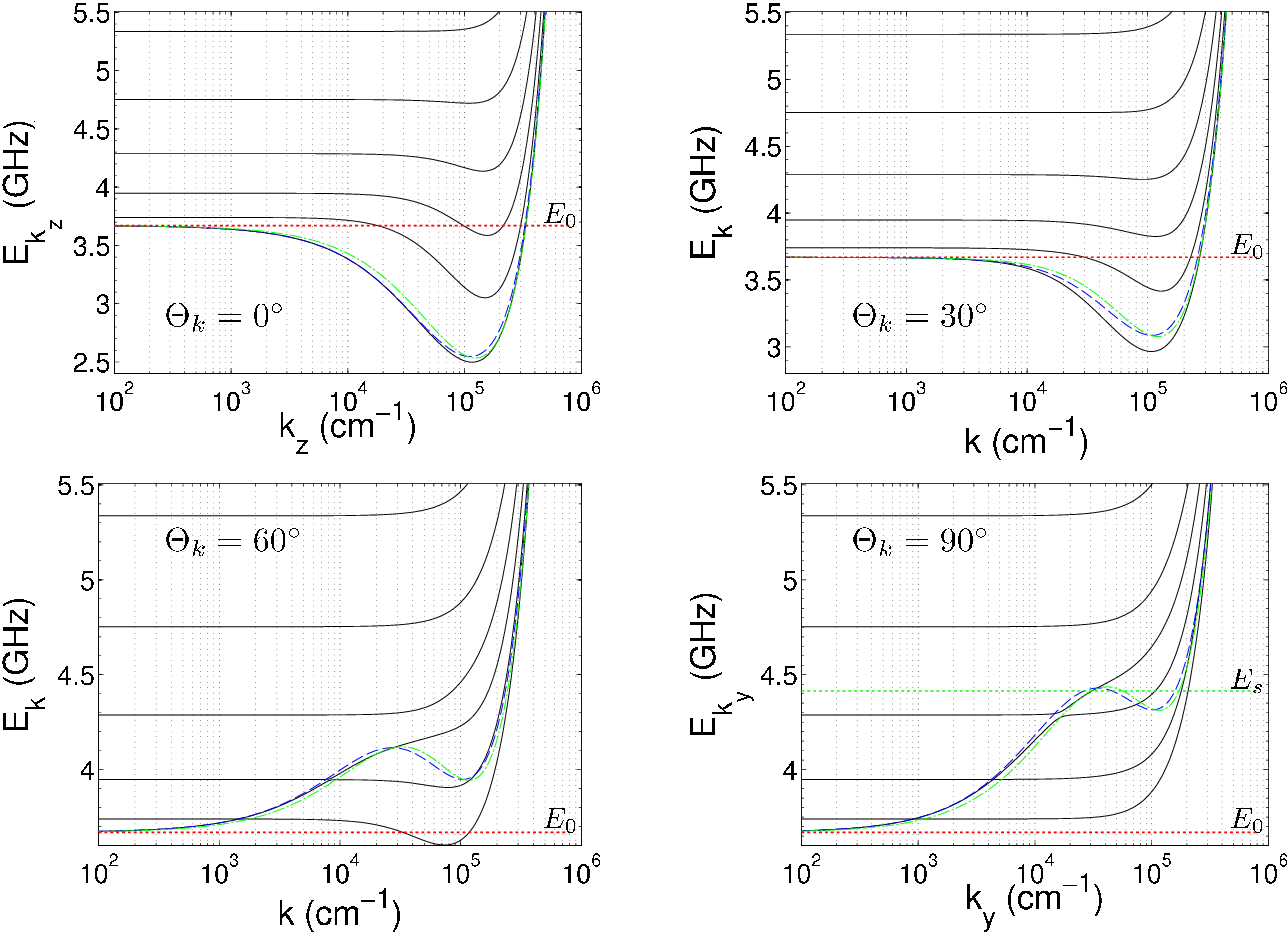}
  \caption{
(Color online)
Spin-wave dispersion of a YIG film with  thickness $d=400a\approx 0.495\,\mu$m in a magnetic field of $700\;\text{Oe}$ for wave-vectors in the plane of the film and
for different propagation 
angles $\Theta_{\bd{k}}=0^\circ\!, 30^\circ\!, 60^\circ\!, 90^\circ$ 
relative to the magnetic field
(from top left to bottom right). The black solid curves represent the results from the numerical approach for the lowest modes. The other curves are obtained from Eq.~(\ref{eq:magnon4app}) using the approximation Eq.~(\ref{eq:form1}) (dashed) and Eq.~(\ref{eq:form2}) (dash dotted) for the form factor $f_{\bd{k}}$. 
The thick dotted line labelled $E_0$ marks the energy of the ferromagnetic resonance 
given by Eq.~(\ref{eq:fm}); the thick dotted line labelled $E_s$ is the energy $E_s=h+\Delta/2$
of the classical surface mode for $\Theta_{\bd{k}}=90^\circ$ at large wave-vectors,
see Eqs.~(\ref{eq:Delta}, \ref{eq:Es}).}
\label{fig:YIG400}
\end{figure*}

\begin{figure}[tb]
    \includegraphics[width = 70mm]{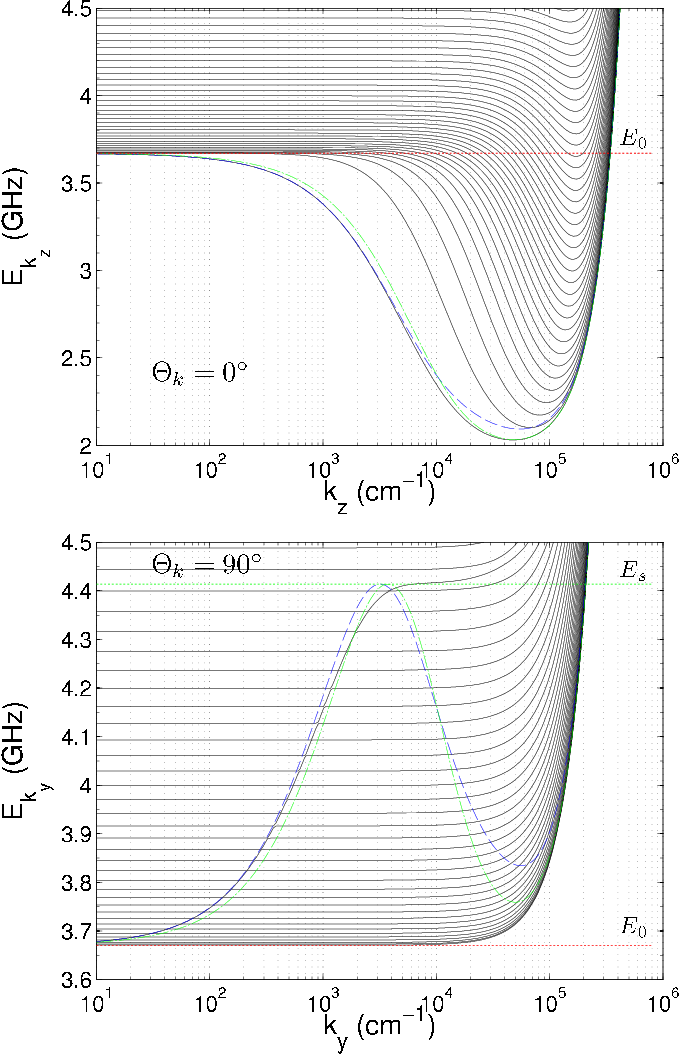}
  \caption{
(Color online)
Spin-wave dispersion of a YIG film with thickness $d=4040a=5\mu\text{m}$ for wave-vectors parallel to the external field $H_e=700\;\text{Oe}$ for $\Theta_{\bd{k}}=0^\circ$ (top) 
and for $\Theta_{\bd{k}}=90^\circ$ (bottom).}
\label{fig:YIG4040}
\end{figure}

If the lattice has $N$ sites in the $x$-direction, then for fixed  two-dimensional wave-vector $\bd{k}$ there are $N$ allowed magnon energies $E_{ n \bd{k}}$, $n =0, \ldots , N -1$, 
which are given by the positive zeros of the secular determinant
 \begin{equation}
 {\rm det} 
 \left( \begin{array}{cc} E_{\bd{k}}  - \mathbf{A}_{\bd{k}} &  - \mathbf{B}_{\bd{k}}
 \\   - \mathbf{B}^{\ast}_{\bd{k}} &    - E_{\bd{k}}  - \mathbf{A}_{\bd{k}}  
\end{array} 
\right) =0.
 \label{eq:det}
 \end{equation}
Here, the
$N \times N$-matrices 
$\mathbf{A}_{ \bd{k}}$ and  $\mathbf{B}_{ \bd{k}}$  are defined by
 \begin{subequations}
 \begin{eqnarray}
[ \mathbf{A}_{ \bd{k}} ]_{ ij } & = &  A_{ \bd{k} } (  x_{ij} ) ,
 \end{eqnarray}
\begin{eqnarray}
[ \mathbf{B}_{ \bd{k}} ]_{ ij } & = &  B_{ \bd{k} } (  x_{ij} ) ,
 \end{eqnarray}
 \end{subequations}
where $1 \leq  i , j \leq N$ label now the lattice sites in the $x$-direction. The condition (\ref{eq:det}) follows simply from the fact that the magnon energies can be identified with the poles of the propagators of the Gaussian field theory defined by the quadratic Hamiltonian (\ref{eq:H2Gauss}). Note that for $N =1$ the condition (\ref{eq:det}) correctly reduces to the diagonalization of the Hamiltonian (\ref{eq:H2Gauss}) via Bogoliubov transformation. 
To obtain the complete magnon spectrum of the thin film ferromagnet 
we have to calculate the dipolar matrices 
in Eq.~(\ref{eq:ab}) which leads to the calculation of the following 
dipolar sums for fixed $x_{ij}$,
\begin{eqnarray}
D_{\bd{k}}^{\alpha \beta} (x_{ij}) &=&  {\sum_{\bd{r}_{ij}}}^\prime e^{-i \bd{k}\cdot \bd{r}_{ij}}  D_{ij}^{\alpha \beta} \nonumber \\
& & \hspace{-20mm} 
= - \mu^2  {\sum_{y_{ij},z_{ij}}}^{ \hspace{-1.5mm} \prime} e^{-i(k_y y_{ij}+k_z z_{ij})} \nonumber \\ 
& &\hspace{-15mm}  \times 
\left[\frac{\delta^{\alpha \beta}}{(x_{ij}^2+y_{ij}^2+z_{ij}^2)^{3/2}}  - \frac{3 r_{ij}^\alpha r_{ij}^\beta }{(x_{ij}^2+y_{ij}^2+z_{ij}^2)^{5/2}}  \right], \nonumber \\
\label{eq:dipsum}
\end{eqnarray}
where $\sum^\prime$ excludes the term  $y_{ij}=z_{ij}= 0$ when $x_{ij}=0$. 
As these sums are slowly converging and previously used summation 
techniques~\cite{Costa00} are not very efficient for small wave-vectors 
needed for the dipolar-dominated spectrum of YIG, 
we use the Ewald summation technique to
obtain fast convergence.
Details of the calculation can be found in the appendix.
To determine the dispersion of all modes we numerically calculate the roots 
of Eq.~(\ref{eq:det}) which can be easily
done up to a film thickness  $d\approx7\,\mu\text{m}$.
Using the fact that  $\mu/\hbar\approx 2.803\times 10^{-3}\,\text{GHz/Oe}$ 
we plot all results in units of GHz which is most convenient for 
experiments using microwave resonators and antennas to detect magnetic excitations.
Typical magnon dispersions for different angles $\Theta_{\bd{k}}$ between the propagation
direction and the magnetic field are shown in Fig.~\ref{fig:YIG400} for a 
film with thickness $d = 400a$.
Since our approach includes all effects of the exchange and the dipolar interaction 
within linear spin wave theory, it reproduces  semiclassical approximations
based on the Landau-Lifshitz equation at long wave-lengths.
In particular, for $\bd{k}\rightarrow 0$ the dispersion of the lowest mode
approaches, independently of the thickness $d$, the classical ferromagnetic resonance energy
\begin{equation}
 E_0=\sqrt{h (h+\Delta)}\;, 
\label{eq:fm}
\end{equation}
where we introduced the characteristic magnon energy due to dipolar interactions,
\begin{equation}
\Delta=4\pi\mu M_S.
\label{eq:Delta}
\end{equation}
The energy $E_0$ is indicated as a dotted line in Fig.~\ref{fig:YIG400}.
The spacing between different magnon modes decreases with
increasing film thickness $d$; for example in Fig.~\ref{fig:YIG4040}
we show the magnon spectrum of a film with thickness $d = 4040a = 5\,\mu$m,
which forms already a quasi-continuum for energies close to the ferromagnetic resonance.
Moreover in the regime $dk\ll1$ one observes that the energy of the $n$~\!-~\!th mode deviates from the first mode by an energy $\Delta E_n=\rho_{\text{ex}}\pi^2 n^2/d^2$ which reflects the well known quadratic behaviour of ferromagnetic spin waves in three dimensions.

\begin{figure}[tb]
   \includegraphics[width = 72mm]{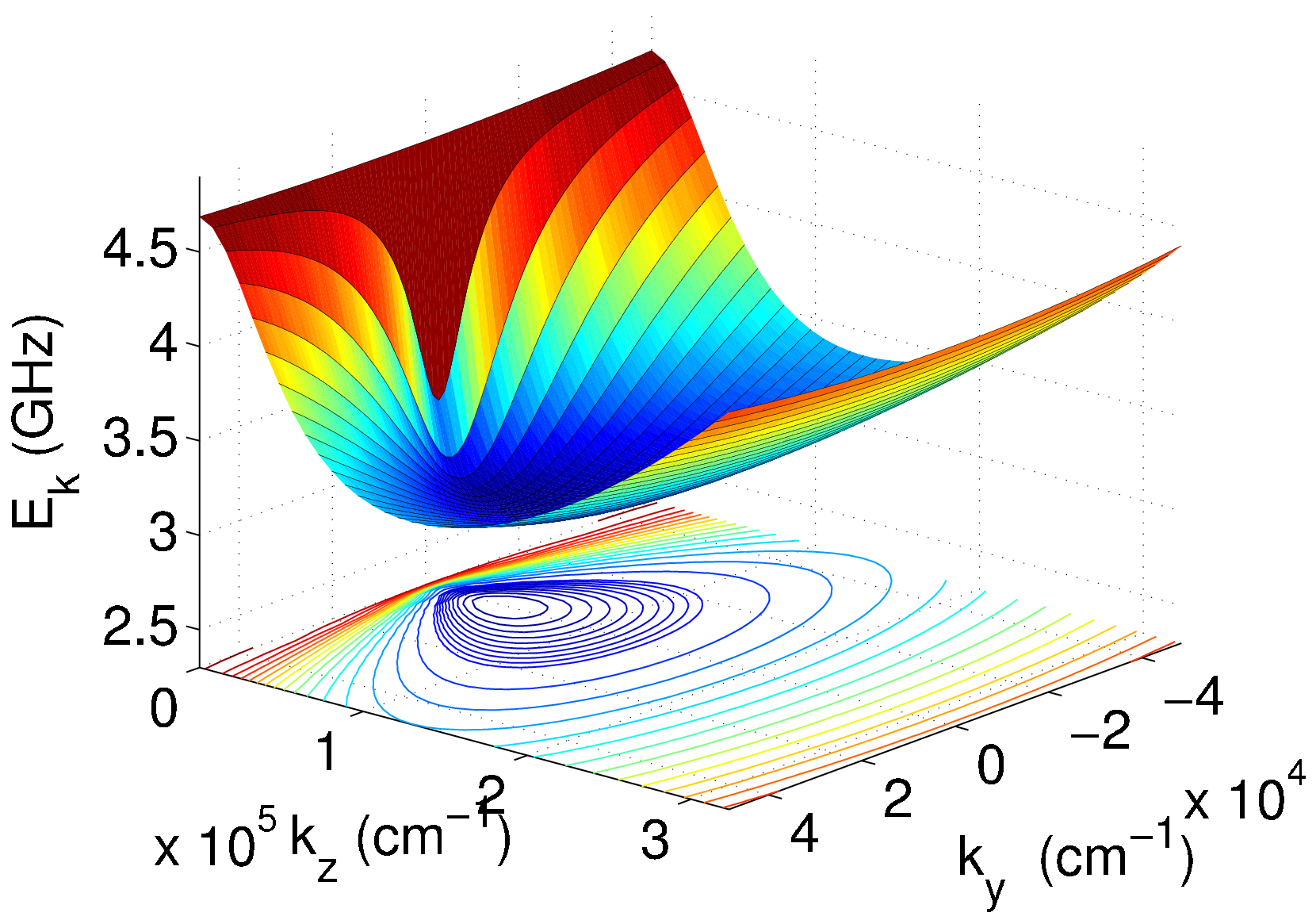}
  \caption{
(Color online) Spin-wave dispersion of the lowest mode of a YIG film with $H_e=1000\;\text{Oe}$ and $d=5.1\mu \text{m}$ obtained from the numerical solution of Eq.~(\ref{eq:det}). Starting from the minimal energy $E_{\text{min}}$ contour lines with the spacing 20 MHz are shown to illustrate the rather flat minimum, whereas for larger energies the distance between contour lines is 100 MHz.}
\label{fig:YIG3d}
\end{figure}

\begin{figure}[tb]
   \includegraphics[width = 75mm]{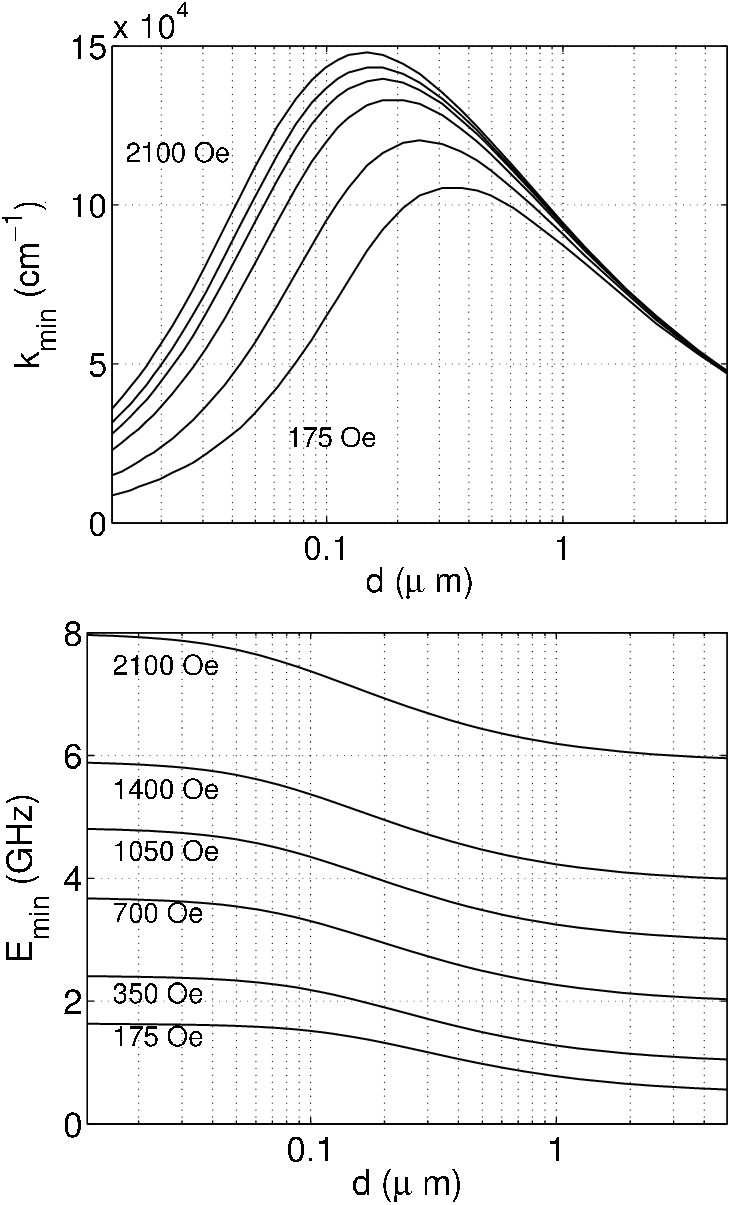}
  \caption{
Momentum $k_{\text{min}}$ and energy $E_{\text{min}}$ of the minimum in the dispersion
of the lowest magnon mode 
for $\Theta_{\bd{k}}=0$ as a function of film thickness for different magnetic fields. From top to bottom $H_e= 2100, 1400, 1050, 700, 350, 175 \;\text{Oe}$.}
\label{fig:min}
\end{figure}

Of particular interest is the lowest magnon mode,
whose dispersion is illustrated in Fig.~\ref{fig:YIG3d}.
For propagation directions parallel to the magnetic field
($\Theta_{\bd{k}}=0$) we recover the well known~\cite{Kalinikos86}
minimum of the lowest magnon mode at a finite wave-vector $\bd{k}_{\text{min}}$
where large magnon densities have been detected 
by their microwave radiance~\cite{Demokritov06,Demokritov08}.
As illustrated in Fig.~\ref{fig:min}, 
the position of the minimum depends on the film thickness as expected 
and is less pronounced for ultrathin films.
With increasing propagation angle
$\Theta_{\bd{k}}$ 
the minimum becomes more shallow
and completely disappears for $\Theta_{\bd{k}} = 90^\circ$.
From Figs.~\ref{fig:YIG400} and \ref{fig:YIG4040} it is also obvious that
for angles $\Theta_{\bd{k}}>45^\circ$ 
the lowest modes shift upwards and tend to hybridize with higher modes.
However, in the absence of additional symmetries
the energy levels never cross.
The magnon modes with  finite 
group velocities $\bd{v}_g( \bd{k})= \bd\nabla_{\bd{k}}E_{\bd{k}}$ tend
upwards and form the quasi-continuous surface mode. 
There are no further hybridizations as soon as the mode energies  reach the 
energy 
 \begin{equation}
E_s=h+ \Delta/ 2
  \label{eq:Es}
 \end{equation}
of the classical surface mode.

\subsection{Uniform mode approximation}
\label{subsec:uniform}
\begin{figure}[tb]
   \includegraphics[width = 65mm]{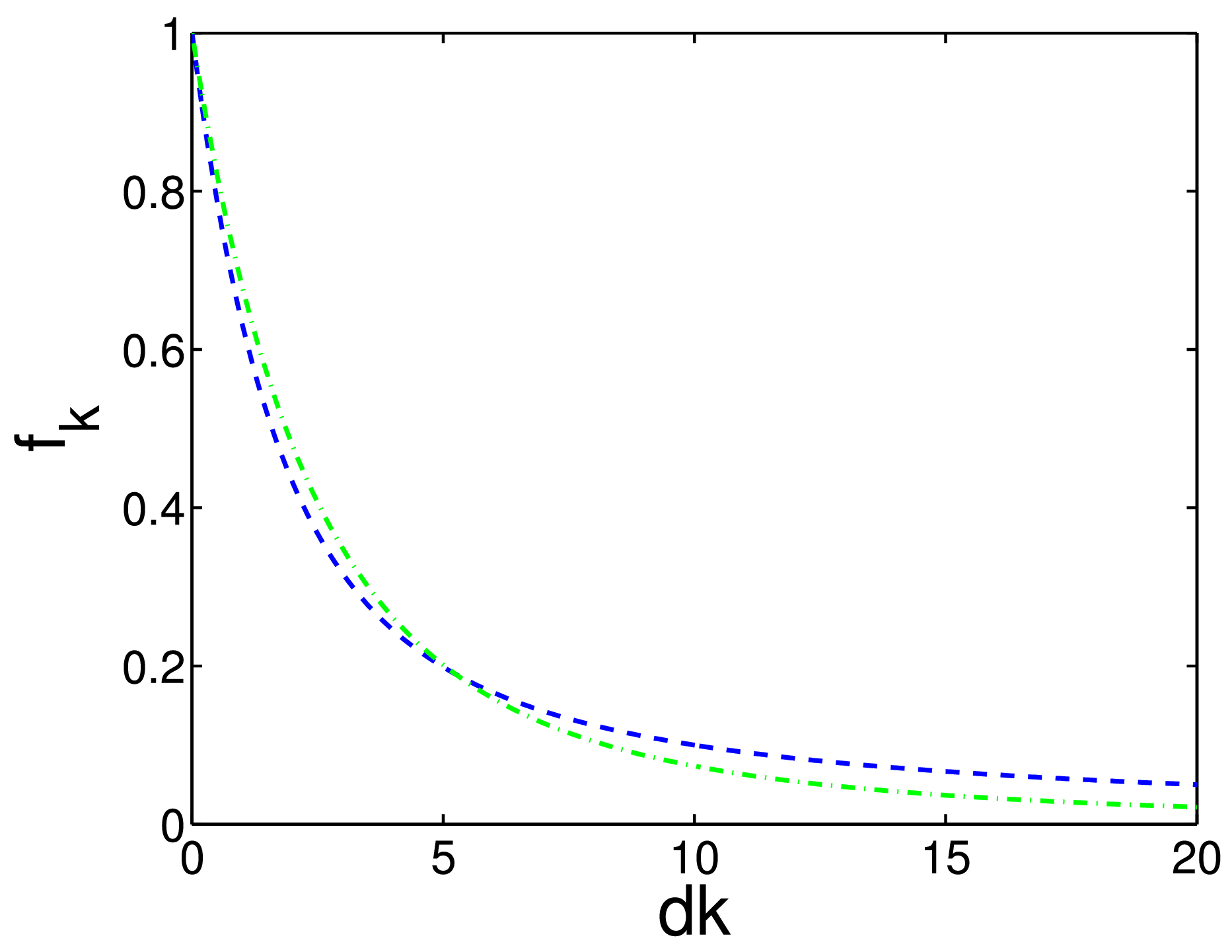}
  \caption{
(Color online) Plot of the form factors Eq.~(\ref{eq:form1}) for the uniform mode approximation (dashed) and Eq.~(\ref{eq:form2}) for the lowest eigenmode approximation (dash dotted). For the samples used in the experiments the minimum of the lowest mode is at $dk\gtrsim 5$ where the lowest eigenmode approximation is more accurate.}
\label{fig:fk}
\end{figure}

The dispersion of the lowest magnon band
can be derived from an effective
in-plane Hamiltonian using various approximations.
In fact, at long wavelengths the dispersion of the
lowest magnon band can also be obtained within the macroscopic
approach based on the Landau-Lifshitz equation~\cite{Kalinikos86}. 
In the simplest approximation, we ignore the fact that  the system is not 
translationally invariant in the $x$-direction  and approximate the corresponding eigenfunctions by plane waves. The lowest magnon band is then obtained by replacing the operators $b_{ \bd{k} } ( x_i )$
in Eq.~(\ref{eq:H2Gauss}) by
 \begin{equation} 
 b_{ \bd{k} } ( x_i ) \approx
 \frac{1}{N} \sum_j  b_{ \bd{k} } ( x_j )  \equiv   \frac{1}{\sqrt{N}}  b_{ \bd{k} }.
 \label{eq:uniform}
 \end{equation}
Then Eq.~(\ref{eq:H2Gauss}) reduces to
  \begin{equation}
 \hat{H}_2^{\rm eff} = \sum_{\bd{k}}
 \left[ A_{\bd{k}} b^{\dagger}_{\bd{k}} b_{\bd{k}} + \frac{ B_{\bd{k}}}{2} b_{ \bd{k}} 
b_{ - \bd{k}}
  + \frac{ B^{\ast}_{\bd{k}}}{2} b^{\dagger}_{ \bd{k}} b^{\dagger}_{ - \bd{k}} \right],
 \label{eq:H2eff}
 \end{equation}
with
 \begin{subequations}
 \begin{eqnarray}
   A_{ \bd{k} } & = & \frac{1}{N} \sum_{ i j}   A_{ \bd{k} } (  x_{ij} )\;, \\
   B_{ \bd{k} } & = & \frac{1}{N} \sum_{ i j}B_{ \bd{k} } (  x_{ij} )\;.
 \end{eqnarray}
\end{subequations}
We parametrize the in-plane wave-vectors as
\begin{equation}
 \bd{k}=| \bd{k} | \left(\cos\Theta_{\bd{k}}\;\bd e_z+\sin\Theta_{\bd{k}}\;\bd e_y\right),
\end{equation}
carry out the integrations over the two infinite directions and obtain for  $x_{ij} \neq 0$
\begin{subequations}
\begin{eqnarray}\label{eq:Dabij1}
 D^{xx}_{\bd{k} } ( x_{ij} )  & = & D_{\bd{k}}(x_{ij}),
 \label{eq:Dxxij}\\
 D^{yy}_{\bd{k} } ( x_{ij} )  & = &-\sin^2\Theta_{\bd{k}} \;D_{\bd{k}}(x_{ij}),
 \label{eq:Dyyij}\\
D^{zz}_{\bd{k} } ( x_{ij} )  & = &-\cos^2\Theta_{\bd{k}}\; D_{\bd{k}}(x_{ij}),
 \label{eq:Dzzij}\\
 D^{xy}_{\bd{k} } ( x_{ij} )  & = &\sin\Theta_{\bd{k}} \;  D_{\bd{k}}(x_{ij}) \Bigl[\frac{|\bd{k}|}{x_{ij}}+\frac{\text{sign}(x_{ij})}{x_{ij}^2}\Bigr], \hspace{7mm}
 \label{eq:Dxyij}\label{eq:Dabij2}
 \end{eqnarray}

\end{subequations}
where
\begin{equation}
 D_{\bd{k}}(x_{ij})=\frac{2 \pi \mu^2}{a^2} |\bd{k} |   e^{ - |\bd{k}|  | x_{ij} | }.
\end{equation}
It should be noted that in the derivation of the dipolar tensor (\ref{eq:Dabij1}-\ref{eq:Dabij2}) we have implicitly assumed that $x_{ij}\neq 0$. This has important consequences when we also replace the sum $\sum_{ x_i}$ by the integral $a^{-1} \int_{ -d/2}^{d/2} dx$. To properly account for the factor $(1-\delta_{ij})$ in the dipole 
tensor in Eq.~(\ref{eq:defdip}) we therefore exclude a sphere of infinitesimal radius 
around $x_{ij}=y_{ij}=z_{ij}=0$ in our integrations, giving rise to the dipole matrix elements
\begin{subequations}
\begin{eqnarray}\label{eq:Dabk1}
 D^{xx}_{\bd{k}} & = &\frac{4 \pi \mu^2}{a^3}\Bigr[ \frac 13 -f_{ \bd{k} }\Bigr], \\
 D^{yy}_{\bd{k}} & = &\frac{4 \pi \mu^2}{a^3}\Bigr[ \frac 13+ \sin^2\Theta_{\bd{k}} 
 \left( f_{  \bd{k}  }-1 \right)\Bigr], \\
 D^{zz}_{\bd{k}} & = &\frac{4 \pi \mu^2}{a^3}\Bigr[ \frac 13+ \cos^2\Theta_{\bd{k}} 
 \left( f_{  \bd{k}  }-1 \right)\Bigr],\label{eq:Dabk2}
 \end{eqnarray}
\end{subequations}\\[0cm]
\noindent
and $D^{xy}_{\bd{k}}=0$. Here we have introduced the form factor~\cite{Tupitsyn08}
 \begin{equation}
  f_{ \bd{k} } = \frac{ 1 - e^{- | \bd{k} |   d}}{ | \bd{k} |  d} = 
1-\frac{ | \bd{k} |  d}{2}+ \mathcal{O} ( \bd{k}^2 d^2 ),
\label{eq:form1}
 \end{equation}
which is shown as a dashed line in Fig.~\ref{fig:fk}.
From Eqs.~(\ref{eq:Dabk1}-\ref{eq:Dabk2}) it is immediately obvious that our dipole elements satisfy the constraint
\begin{equation}
 D_{\bd{k}}^{xx}+D_{\bd{k}}^{yy}+D_{\bd{k}}^{zz}=0 ,
\end{equation}
which follows also directly from the definition (\ref{eq:defdip}) of the dipolar tensor.
In terms of the above dipole matrix elements we can now write the coefficients  
$A_{\bd{k}}$ and $B_{\bd{k}}$ as
\begin{subequations}
\begin{eqnarray}
   A_{ \bd{k} }&=& h + JS\bigl[4-2\cos (k_ya)-2\cos(k_za)\bigr]\nonumber\\
&&-\frac S2 \bigl(D_{\bd{k}}^{xx}+D_{\bd{k}}^{yy}\bigr)+\frac \Delta 3 , \\
B_{ \bd{k} }&=&-\frac S2 \bigl(D_{\bd{k}}^{xx}-D_{\bd{k}}^{yy}\bigr)\;.
\end{eqnarray}
\end{subequations}
The magnon dispersion $E_{\bd{k}}$  is now obtained by diagonalizing the effective Hamiltonian (\ref{eq:H2eff}) via a Bogoliubov transformation~\cite{Akhiezer68,Mattis06}, resulting in
 \begin{equation}
  E_{\bd{k}} = \sqrt{ A_{\bd{k}}^2 - | B_{\bd{k} } |^2 }=\sqrt{( A_{\bd{k}}-  |B_{\bd{k} }|)( A_{\bd{k}}+  |B_{\bd{k} }|) }.
 \label{eq:magnon4}
 \end{equation}
If we expand the terms involving the exchange interaction for small $\bd{k}$ (which is sufficient for the energy range in experiments on thin films) we get
\begin{equation}
 JS\bigl[4-2\cos (k_ya)-2\cos(k_za)\bigr]\approx JSa^2 \bd{k}^2=\rho_{\rm ex} \bd{k}^2.
\end{equation}
We can then simplify the dispersion to
\begin{eqnarray}
 E_{\bd{k}} & =& 
 \nonumber
 \\
 & & 
\hspace{-12mm} 
\sqrt{[h +\rho_{\rm ex} \bd{k}^2+\Delta (1-f_{ \bd{k}} ) \sin^2\Theta_{\bd{k}}][h +
\rho_{\rm ex} \bd{k}^2+\Delta f_{\bd{k}}]}, 
\nonumber
 \\
 & &
\label{eq:magnon4app}
\end{eqnarray}
which is indicated by dashed lines in Figs.~\ref{fig:YIG400}, \ref{fig:YIG4040} 
and \ref{fig:zoom}. Obviously, this approximation  is 
in good qualitative agreement with the numerical result for the lowest mode 
as long as $\Theta_{\bd{k}}\lesssim 45^\circ$. 
Eq.~ (\ref{eq:magnon4app}) has been used  in 
Ref.~\cite{Tupitsyn08} to discuss
the role of magnon-magnon interactions in YIG and
has  recently been rederived by
Rezende~\cite{Rezende09}.
Note, however, that in Fig.~\ref{fig:YIG4040}
one clearly sees deviations from the numerical result at 
intermediate wave-vectors in the interval $5\lesssim |\bd{k}|d \lesssim 50$. 
By comparing Fig.~\ref{fig:YIG400} with  Fig.~\ref{fig:YIG4040} we conclude that
for the samples used in experiments~\cite{Demokritov06,Demidov07,Dzyapko07,Demidov08,Demokritov08,Serga07,Schaefer08,Chumak09,Neumann08}
the minimum of the dispersion is exactly in this range
so that better analytical approximations are needed for a more
accurate description of the magnon dispersion in the vicinity of the minimum.

\subsection{Lowest eigenmode approximation}
\label{subsec:eigen}
\begin{figure}[tb]
   \includegraphics{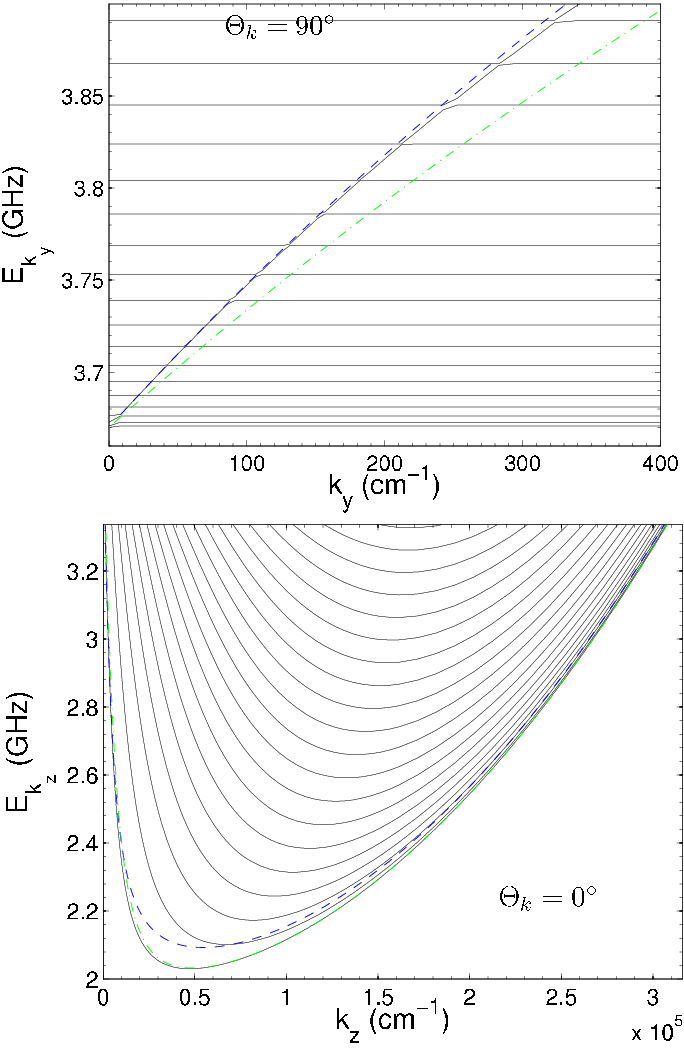}
  \caption{
(Color online) 
Enlarged details of Fig.~\ref{fig:YIG4040} on a linear momentum scale:
spin-wave dispersion of a YIG film with thickness 
$d=4040a=5\mu\text{m}$ for wave-vectors parallel to the external magnetic
field $H_e=700\;\text{Oe}$ ($\Theta_{\bd{k}}=0^\circ$, top) and for 
wave-vectors perpendicular to the magnetic field
($\Theta_{\bd{k}}=90^\circ$, bottom).
The solid lines are exact numerical results obtained from the solution
of Eq.~(\ref{eq:det}). The  dashed line
is our approximate expression (\ref{eq:magnon4app}) for the
lowest magnon band, using the uniform mode approximation
(\ref{eq:form1}) for the form factor. The dashed-dotted line
is the lowest magnon band with form factor (\ref{eq:form2}) given by the
lowest eigenmode approximation.
}
\label{fig:zoom}
\end{figure}

\begin{figure}[tb]
   \includegraphics{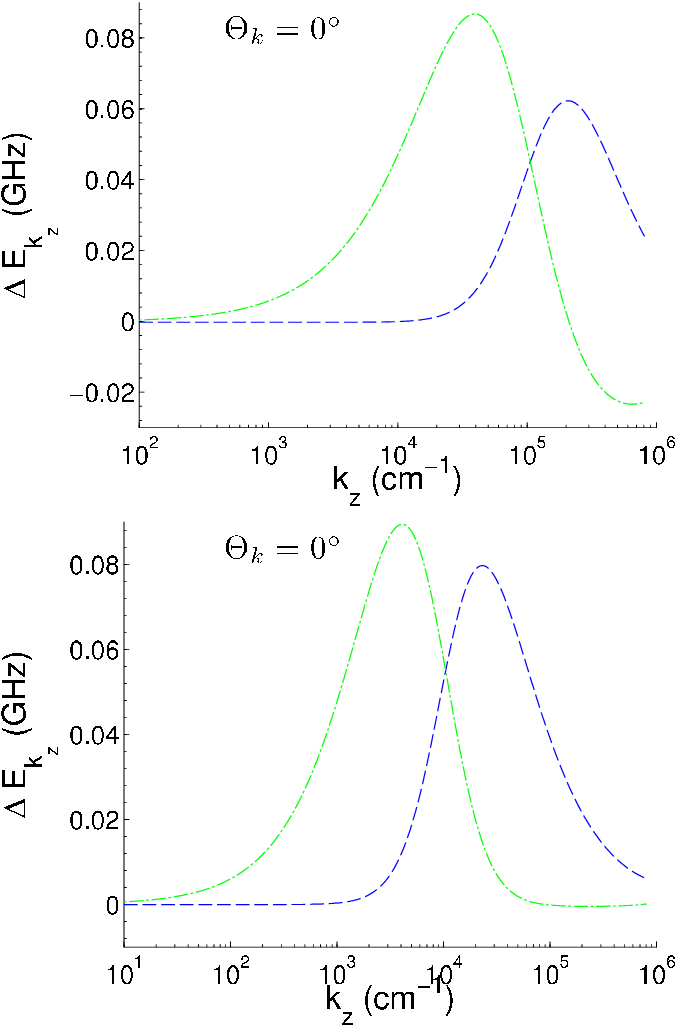}
  \caption{
(Color online) 
Accuracy of our two analytical approximations for the lowest magnon mode for a 
YIG film with $N=400$ layers (top) and $N=4040$ layers (bottom).
Dashed line: difference  between the numerical result and the uniform mode approximation;
dash-dotted line: difference  between the numerical result and
the lowest eigenmode approximation.}
\label{fig:compare}
\end{figure}

To derive the dispersion of the lowest magnon mode more systematically,
suppose that the $\psi_{ n \bd{k}} ( x_i )$ form (for fixed $\bd{k}$) a complete set of 
orthogonal functions with respect to the $x$-direction, i.e.,
 \begin{eqnarray}
 \sum_{ x_i} \psi^{\ast}_{ n \bd{k}} ( x_i ) \psi_{ m \bd{k}} ( x_i )
 & = & \delta_{ n m} ,
 \\
  \sum_{ n} \psi_{ n \bd{k}} ( x_i ) \psi^{\ast}_{ n \bd{k}} ( x_{j} )
 & = & \delta_{ i j} .
\end{eqnarray}
We may then expand the operators $b_{\bd{k}} ( x_i )$ in this basis,
 \begin{equation}
b_{\bd{k}} ( x_i ) = \sum_n \psi_{ n  \bd{k}} ( x_i ) b_{ n \bd{k}},
 \label{eq:bexpansion} 
\end{equation}
where
 \begin{equation}
 b_{ n \bd{k}}  = \sum_{ x_i}  \psi^{\ast}_{ n  \bd{k}} ( x_i )  b_{\bd{k}} ( x_i ).
 \end{equation}
Let us retain in  the expansion (\ref{eq:bexpansion}) only the $n=0$ term,
  \begin{equation}
b_{\bd{k}} ( x_i )  \approx  \psi_{ 0  \bd{k}} ( x_i ) b_{ 0 \bd{k}},
 \label{eq:b0approx} 
\end{equation}
If we choose  $\psi_{ 0  \bd{k}} ( x_i ) = 1/ \sqrt{N}$ (corresponding to the $n=0$ term
in the plane wave expansion) and
identify $ b_{ \bd{k}} \equiv b_{ 0 \bd{k}}$ we recover Eq.~(\ref{eq:uniform}).
However, a truncated expansion in plane waves seems not to be a good approximation 
for a system of finite width.
To improve on the approximation (\ref{eq:uniform}) it is better to
expand in terms of the eigenfunctions of the exchange matrix given in Eq.~(\ref{eq:exmatrix}),
\begin{equation}
 \sum_{ x_j} J_{\bd{k}} ( x_{ij} ) \psi_{ n \bd{k}  } ( x_j ) = \lambda_{ n \bd{k}  } 
\psi_{ n \bd{k}  } ( x_i ).
 \end{equation}
For open boundary conditions these are standing waves with nodes at $x = \pm d/2$, i.e.,
 \begin{equation}
\psi_{ n \bd{k} } ( x_i ) = \sqrt{ \frac{2}{N}} \sin [ k_n (x_i+d/2 )] ,
 \end{equation}
where $k_n = (n+1) \pi /d$, $n=0,1,\ldots, N-1$.
The approximation (\ref{eq:b0approx}) then reduces to
 \begin{eqnarray} 
 b_{ \bd{k} } ( x_i ) & \approx &  \sqrt{ \frac{2}{N}} \cos ( k_0 x_i )  b_{  \bd{k} } ,
 \label{eq:openboundary}
 \end{eqnarray}
with $k_0 = \pi /d$, and
 \begin{equation}
b_{  \bd{k} } =
  \sum_{x_i}  \sqrt{ \frac{2}{N}} \cos ( k_0 x_i )     b_{ \bd{k} } ( x_i  ) .
 \end{equation}
Substituting Eq.~(\ref{eq:openboundary}) into 
Eq.~(\ref{eq:H2Gauss}) we obtain again an effective Hamiltonian of the form 
(\ref{eq:H2eff}), but now with
 \begin{subequations}
  \begin{eqnarray}
   A_{ \bd{k} } & = & \frac{2}{N} \sum_{ i j}   
\cos ( k_0 x_i ) \cos ( k_0 x_j ) A_{ \bd{k} } (  x_{ij} )    , \hspace{7mm}
 \label{eq:Akopen}
 \\
B_{ \bd{k} } & = & \frac{2}{N} \sum_{ ij}
\cos ( k_0 x_i ) \cos ( k_0 x_j )
B_{ \bd{k} }(x_{ij}) .
 \label{eq:Bkopen}
 \end{eqnarray}
 \end{subequations}
Again the summations are replaced by integrations which can be carried out analytically and result in the dispersion of the form of Eq.~(\ref{eq:magnon4app}), but with the form factor now given by
\begin{eqnarray}
 f_{\bd{k}} &=&1-|\bd{k} d|\frac{| \bd{k} d|^3+| \bd{k} d|\pi^2+2\pi^2(1+
 e^{-|\bd{k} d|})}{(\bd{k}^2d^2+\pi^2)^2}\nonumber\\
& = &1-\frac 4{\pi^2} |\bd{k} | d+\mathcal{O} ( \bd{k}^2 d^2 ).
\label{eq:form2}
\end{eqnarray}
In Fig.~\ref{fig:fk} we compare this form factor
with the corresponding form factor (\ref{eq:form1}) obtained within the
uniform mode approximation.
Obviously, there are only small differences: 
the linear coefficient in the Taylor series is different, resulting  
in a smaller slope of the dispersion (\ref{eq:magnon4app}) at $\bd{k}=0$. 
To estimate the validity of these approximations,
we compare them in Figs.~\ref{fig:zoom} and \ref{fig:compare}
with our numerically exact results of the model (\ref{eq:hamiltonian})  for experimentally relevant parameters.
In Fig.~\ref{fig:zoom} we show the relevant details of
 Fig.~\ref{fig:YIG4040} on a linear momentum scale; obviously for
sufficiently large wave-vectors
 the lowest eigenmode approximation  is more accurate and
practically lies on top of the numerically exact result. On the other hand, for
small wave-vectors the uniform mode approximation 
fits better, as illustrated by the lower part of Fig.~\ref{fig:zoom}.
A more quantitative comparison between these two approximations
for wave-vectors parallel to the magnetic field and
different film thickness is shown in Fig.~\ref{fig:compare}.
Roughly, for  $d | \bd{k} | \lesssim 5$ the uniform mode approximation 
fits better, while for 
for $d | \bd{k} | \gtrsim 5$ the lowest eigenmode approximation 
gives a better agreement with the numerical results.
As there are not many states at small wave-vectors, 
we believe that the lowest eigenmode approximation 
is more suitable for a quantitative 
description of the magnon dispersion, in particular
if one is interested in physical effects 
related to the dispersion minimum at $\bd{k}_\text{min}$. 
On the other hand, both analytical approximations
describe the quasi-continuous surface mode for large $\Theta_{\bd{k}}> 45^\circ$ 
rather accurately as long as the wave-vectors are sufficiently small so that
the dispersion curves tend upwards due to the presence of exchange interaction~\cite{Cottam94} and hybridize.

\section{Conclusions and outlook}
\label{sec:conclusions}

In this work we have presented a comprehensive discussion of the
spin-wave spectra of experimentally relevant samples of thin films 
of the magnetic insulator YIG. Starting from an effective spin-$S$ Heisenberg Hamiltonian
with exchange and dipole-dipole interactions on a cubic lattice, we have used
a truncated Holstein-Primakoff transformation to obtain an effective quadratic
boson Hamiltonian whose eigen-energies can be identified with the magnon energies.
We have then used numerical methods to calculate the magnon dispersions 
for experimentally relevant films  
with thickness corresponding to a few thousand lattice spacings
without further approximations.
 In order to carry out the dipolar sums entering the secular determinant,
the use of  efficient Ewald summation techniques was necessary.
We have also estimated the accuracy of two different analytical 
approximations for the lowest magnon  band: the uniform mode approximation (where
the transverse spatial variation of the eigenmodes is ignored) and the
lowest eigenmode approximation (where the lowest transverse mode is approximated by the
lowest eigenmode of the exchange matrix). For realistic films, the latter approximation is more 
accurate for wave-vectors in the vicinity of the dispersion  minimum.

 In contrast to the phenomenological approach based on the
Landau-Lifshitz equation~\cite{Kalinikos86,Cottam94}, in  our microscopic approach
it is straightforward to systematically take into account interaction effects. 
In future work we shall carefully
derive the momentum-dependent 
interaction vertices of the bosonized effective in-plane Hamiltonian using the 
lowest eigenmode approximation, which 
according to our investigations in Sec.~\ref{sec:spectra}
is more accurate in the
experimentally interesting regime of wave-vectors close to $\bf{k}_{\rm min}$.
Although rough estimates of the interactions vertices can be found
in the literature~\cite{Tupitsyn08}, more accurate microscopic calculations 
which properly 
take into account the momentum-dependence of the vertices are needed in order
to gain a better understanding of the role of
spin-wave interactions in the experiments~\cite{Demokritov06,Demidov07,Dzyapko07,Demidov08,Demokritov08,Serga07,Schaefer08,Chumak09,Neumann08}.
For example, it would be interesting to know  the intrinsic damping of the lowest
magnon mode  for wave-vectors close to $\bd{k}_{\rm min}$ due to 
magnon-magnon interactions.

\begin{acknowledgement}
We would like to thank A. A. Serga, T. Neumann and P. Priooznia for helpful discussions and
gratefully acknowledge financial support by the
SFB/TRR49 and the DAAD/PROBRAL-program.
\end{acknowledgement}

\begin{appendix}
\renewcommand{\theequation}{A.\arabic{equation}}
\section*{APPENDIX: DIPOLAR SUMS}
\setcounter{equation}{0}

For the calculation of the slowly converging sums in Eq.~(\ref{eq:dipsum}) we introduce the quantity
\begin{equation}
I_{\bd{k}} (x_{ij}) = \mu^2 \sum_{y_{ij}, z_{ij}} \frac{e^{-i(k_y y_{ij}+k_z z_{ij})}}{(x_{ij}^2+y_{ij}^2+z_{ij}^2)^{5/2}},
\end{equation}
and get the dipolar sums as derivatives,
\begin{subequations}
\begin{eqnarray}
 \hspace{-5mm} D_{\bd{k}}^{yy} (x_{ij}) &=&  \left[ \frac{\partial^2}{\partial k_z^2} - 2 \frac{\partial^2}{\partial k_y^2} - x_{ij}^2 \right] I_{\bd{k}} (x_{ij})\;,\\ 
 \hspace{-5mm}D_{\bd{k}}^{zz} (x_{ij}) &=&  \left[ \frac{\partial^2}{\partial k_y^2} - 2 \frac{\partial^2}{\partial k_z^2} - x_{ij}^2 \right] I_{\bd{k}} (x_{ij})\;, \\ 
 \hspace{-5mm}D_{\bd{k}}^{xx} (x_{ij}) &=&  \left[ \frac{\partial^2}{\partial k_y^2} + \frac{\partial^2}{\partial k_z^2} + 2x_{ij}^2 \right] I_{\bd{k}} (x_{ij})\;,\\ 
 \hspace{-5mm}D_{\bd{k}}^{xy} (x_{ij}) &=&   i \ 3  x_{ij} \frac{\partial}{\partial k_y}  I_{\bd{k}} (x_{ij}).
\end{eqnarray} 
\label{eq:dab}
\end{subequations}\\[0cm]
\noindent
Note the symmetry $D_{\bd{k}}^{xx}=D_{\tilde{\bd{k}}}^{xx}$ and the relation $D_{\bd{k}}^{yy}=D_{\tilde{\bd{k}}}^{zz}$ with $\tilde{\bd{k}}=k_z\bd e_y+k_y\bd e_z$.
To evaluate the above expressions, 
we consider the cases  $x_{ij}\not= 0$ and $x_{ij} = 0$ separately.

\section{Case $x_{ij}\not= 0$}

Using the identity
\begin{equation}
\int_0^\infty x^{n-1/2} e^{-\alpha x} dx = \sqrt{\pi} 2^{-n} \alpha^{-n-1/2} (2n-1)!!,
\label{frazione}
\end{equation}
where $n>0, \; \textrm{Re }\alpha >0$, and introducing the dummy variable $\varepsilon$ to check the results, we can rewrite
\begin{eqnarray}
I_{\bd{k}} (x_{ij}) &=& \mu^2 \frac{4}{3} \sqrt{\frac{ \varepsilon^5}{\pi}} \int_0^\infty dt\ t^{3/2} e^{-x_{ij}^2\varepsilon  t} \nonumber \\ &\times& {\sum_{y_{ij} z_{ij}}} e^{-(y_{ij}^2+z_{ij}^2) \varepsilon t } e^{-i(k_y y_{ij}+ k_z z_{ij})}.
\label{unsplited}
\end{eqnarray}
We split the integral in two parts and use Ewald's method~\cite{Ziman79,Bartosch06}
\begin{equation}
\sum_{\bd{r}} e^{-\varepsilon t |\bd{r}|^2} e^{-i \bd{k} \cdot \bd{r}} = \frac{\pi}{a^2 \varepsilon t} \sum_{ \bd{g}} e^{-\frac{|\bd{k} + \bd{g}|^2}{4 \varepsilon t} },
\label{Ewald}
\end{equation}
to transform the lattice sum into a lattice sum in reciprocal space, where $\bd{g} = g_y {\bd e}_y +  g_z {\bd e}_z$ is a reciprocal lattice vector. This yields
\begin{eqnarray}
I_{\bd{k}} (x_{ij}) &=& \frac{4}{3} \mu^2 \left[  \frac{\sqrt{\pi \varepsilon^3}}{a^2}  \sum_{\bd{g}} \int_0^1 dt\ t^{1/2} e^{-\frac{|\bd{k} + \bd{g}|^2}{4\varepsilon t}- x_{ij}^2 \varepsilon t} \right. \nonumber \\ 
&&  \hspace{7mm} + \left.  \sqrt{\frac{ \varepsilon^5}{\pi}}   \sum_{\bd{r}} e^{- i \bd{k} \cdot \bd{r}} \varphi_{3/2}(|{\bd r_{ij}}|^2 \varepsilon) \right],
\label{eq:ik}
\end{eqnarray}
where  
 \begin{equation}
\varphi_\nu (z) = \int_1^\infty dt \ t^\nu e^{-zt}
 \end{equation}
 is the Misra function,\footnote{The Misra function
is defined in terms of the exponential integral $\textrm{Ei}_{\nu}(z) = 
\int_1^\infty dx e^{-xz} x^{-\nu}$ via
 $\varphi_\nu (z) = \textrm{ Ei}_{-\nu}(z)$.}
which for  $\nu = 3/2$ and $z \equiv x > 0$ can also be written as
\begin{equation}
 \varphi_{3/2}(x)=e^{-x}\frac{3+2x}{2x^2}
+\frac{3\sqrt \pi \text{Erfc}(\sqrt x)}{4x^{5/2}}.
 \end{equation}
Evaluating the integral in Eq.~(\ref{eq:ik}) for  $p>0$ and $q^2>0$ using
\begin{eqnarray}
\int_0^1 &dt&  t^{1/2} e^{-q^2 t} e^{-p^2/t} = - \frac{ e^{-p^2-q^2}  }{q^2}  \nonumber \\
&+&  \frac{\sqrt{\pi}}{4q^3}  \left[ e^{-2pq} (1+2pq) \textrm{ Erfc}(p-q) \right. \nonumber \\
 && \hspace{5mm} + \left.  e^{2pq} (-1+2pq) \textrm{ Erfc}(p+q) \right],
\end{eqnarray}
we can calculate the derivatives according to Eqs.~(\ref{eq:dab}) and finally get the result for the sums of the dipole interactions between spins located in different atomic layers,
\begin{subequations}
\begin{eqnarray}
D_{\bd{k}}^{xx} (x_{ij}) &=&-\frac{\pi\mu^2}{a^2}\sum_{\bd{g}}\Bigl\{\frac{8\sqrt \varepsilon}{3\sqrt \pi} e^{-p^2-q^2}
-|\bd{k}+\bd{g}|f(p,q)\Bigr\}
\notag\\ &&\hspace{-2cm}
-\frac{4\varepsilon^{5/2}\mu^2}{3\sqrt \pi}\sum_{\bd{r}}(\bd r_{ij}^2-3x_{ij}^2)\cos(k_yy_{ij})\cos(k_zz_{ij})\varphi_{3/2}(\bd r_{ij}^2 \varepsilon)\;,\notag\\
\\
D_{\bd{k}}^{yy} (x_{ij}) &=&\frac{\pi\mu^2}{a^2}\sum_{\bd{g}}\Bigl\{\frac{4\sqrt \varepsilon}{3\sqrt \pi} e^{-p^2-q^2}
-\frac{(k_y+g_y)^2}{|\bd{k}+\bd{g}|} f(p,q)\Bigr\}
\notag\\&&\hspace{-2cm}
-\frac{4\varepsilon^{5/2}\mu^2}{3\sqrt \pi}\sum_{\bd{r}}(\bd r_{ij}^2-3y_{ij}^2)\cos(k_yy_{ij})\cos(k_zz_{ij})\varphi_{3/2}(\bd r_{ij}^2 \varepsilon)\;,\notag\\
\\
D_{\bd{k}}^{xy} (x_{ij}) &=&i\frac{\pi\mu^2}{a^2}\text{sig}(x_{ij})\sum_{\bd{g}}(k_y+g_y) f(p,q)
\notag\\&&\hspace{-2cm}
+i\frac{4\varepsilon^{5/2}\mu^2}{\sqrt \pi} x_{ij} \sum_{\bd{r}} y_{ij}\sin(k_yy_{ij})\cos(k_zz_{ij}) \varphi_{3/2}(\bd r_{ij}^2\varepsilon)\;,\notag\\
\end{eqnarray}
\label{eq:sumdab}
\end{subequations}\\[0cm]
\noindent
where we have used the abbreviations $q=x_{ij}\sqrt\varepsilon$ and $p=|\bd{k}+\bd{g}|/(2\sqrt \varepsilon)$ and have introduced the function
\begin{equation}
 f(p,q)= e^{-2pq}\text{Erfc}(p-q)+e^{2pq}\text{Erfc}(p+q).
\end{equation}
For the simple cubic lattice the components of the reciprocal lattice vectors are $g_y = 2\pi m,\ g_z = 2\pi n, \ \{m,n\} \in {\mathbb{Z}}$. Note that this result is independent of the variable $\varepsilon$ which shifts the weight from the sums in real space to the sums in reciprocal space as $\varepsilon\rightarrow 0$ and therefore simplifies to the sums given in
Ref.~\cite{Costa00} for $\varepsilon=0$.

\section{Case $x_{ij} = 0$}
In this case we have to exclude  the point $\bd{r} = 0$ because there is no self interaction. We therefore introduce the dummy variable $\bd{x} = y {\bd e}_y + z  {\bd e}_z $,
\begin{equation}
I_{\bd{k}} (x_{ij}\equiv 0) = I_{\bd{k}} = \sum_{\bd{r}} \frac{e^{-i \bd{k} \cdot \bd{r}}}{|\bd{r} - \bd{x}|^5}.
\label{eq:ixnull}
\end{equation}
Using Eq.~(\ref{frazione}) we can rewrite Eq.~(\ref{eq:ixnull}) and use the 
transformation on the reciprocal lattice
\begin{equation}
\sum_{\bd{r}} e^{- i \bd{k} \cdot \bd{r} - |\bd{x} - \bd{r}|^2 \varepsilon t} = \frac{\pi}{a^2 \varepsilon t} \sum_{\bd{g}} e^{i (\bd{g} + \bd{k})\cdot \bd{x} - \frac{|\bd{g} + \bd{k}|^2}{4\varepsilon t}}.
\end{equation}
We subtract $1/|\bd{x}|^5$ from the sum on the left, which is equivalent to removing this first term in the sum over $\bd{r}$ in our dipole sum. In the limit $\bd{x} \rightarrow 0$ we obtain,
\begin{eqnarray}
 I_{\bd{k}}&=&\frac{8\varepsilon^{3/2}\sqrt \pi}{9 a^2}\sum_{\bd{g}} \bigl[e^{-p^2}(1-p^2)+2\sqrt \pi p^3 \text{Erfc}(p)\bigr]\notag\\
&&\hspace{-10mm} +\frac{8\varepsilon^{5/2}}{9\sqrt \pi}{\sum_{\bd{r}}}^{\prime}\bigl[e^{-{\bd{r}}^2 \varepsilon}(1-2{\bd{r}}^2\varepsilon)
+2\sqrt \pi |\bd{r}| \sqrt\varepsilon \text{Erfc}(|\bd{r}|\sqrt \varepsilon)\bigr]
 \nonumber
 \\
& & \hspace{-10mm}
-\frac{8\varepsilon^{5/2}}{15 \pi}.
\end{eqnarray}
After taking the derivatives according Eq.~(\ref{eq:dab}) we note that we get for the dipolar sums the limit $q=x_{ji}\varepsilon\rightarrow0$ of Eq.~(\ref{eq:sumdab}) and therefore the removing of the origin does not make any difference in the calculation of the dipol sums except for omitting the term $\bd{r}=0$ in the real space sums.
\end{appendix}


\begin{thebibliography}{10}
\providecommand{\url}[1]{{\tt #1}}
\providecommand{\urlprefix}{URL }
\providecommand{\eprint}[2][]{\url{#2}}

\bibitem{Demokritov06}
S.~O. Demokritov, V.~E. Demidov, O.~Dzyapko, G.~A. Melkov, A.~A. Serga,
  B.~Hillebrands, and A.~N. Slavin, Nature {\bf 443}, 430 (2006)

\bibitem{Demidov07}
V.~E. Demidov, O.~Dzyapko, S.~O. Demokritov, G.~A. Melkov, and A.~N. Slavin,
  Phys. Rev. Lett. {\bf 99}, 037205 (2007)

\bibitem{Dzyapko07}
O.~Dzyapko, V.~E. Demidov, S.~O. Demokritov, G.~A. Melkov, and A.~N. Slavin,
  New J. Phys. {\bf 9}, 64 (2007)

\bibitem{Demidov08}
V.~E. Demidov, O.~Dzyapko, S.~O. Demokritov, G.~A. Melkov, and A.~N. Slavin,
  Phys. Rev. Lett. {\bf 100}, 047205 (2008)

\bibitem{Demokritov08}
S.~O. Demokritov, V.~E. Demidov, O.~Dzyapko, G.~A. Melkov, and A.~N. Slavin,
  New J. Phys. {\bf 10}, 045029 (2008)

\bibitem{Kalinikos86}
B.~A. Kalinikos and A.~N. Slavin, J. Phys. C: Solid State Phys. {\bf 19}, 7013
  (1986)

\bibitem{Cottam94}
M.~G. Cottam, editor, {\em Linear and Nonlinear Spin Waves in Magnetic Films
  and Superlattices\/}, World Scientific, Singapore (1994)

\bibitem{Serga07}
A.~A. Serga, A.~V. Chumak, A.~Andre, G.~A. Melkov, A.~N. Slavin, S.~O.
  Demokritov, and B.~Hillebrands, Phys. Rev. Lett. {\bf 99}, 227202 (2007)

\bibitem{Schaefer08}
S.~Sch{\"a}fer, A.~V. Chumak, A.~A. Serga, G.~A. Melkov, and B.~Hillebrands,
  Appl. Phys. Lett. {\bf 92}, 162514 (2008)

\bibitem{Chumak09}
A.~V. Chumak, A.~A. Serga, B.~Hillebrands, G.~A. Melkov, V.~Tiberkevich, and
  A.~N. Slavin, Phys. Rev. B {\bf 79}, 014405 (2009)

\bibitem{Neumann08}
T.~Neumann, A.~A. Serga, and B.~Hillebrands, Appl. Phys. Lett. {\bf 93}, 252501
  (2008)

\bibitem{Holstein40}
T.~Holstein and H.~Primakoff, Phys. Rev. {\bf 58}, 1098 (1940)

\bibitem{Dyson56}
F.~J. Dyson, Phys. Rev. {\bf 102}, 1217, 1230 (1956)

\bibitem{Maleev57}
S.~Maleev, Zh. Eksper. Teor. Fiz. {\bf 33}, 1010 (1957)

\bibitem{Akhiezer68}
A.~I. Akhiezer, V.~G. Bar'yakhtar, and S.~V. Peletminskii, {\em Spin Waves\/},
  North Holland, Amsterdam (1968)

\bibitem{Mattis06}
D.~C. Mattis, {\em The theory of magnetism made simple\/}, World Scientific,
  Singapore (2006)

\bibitem{Erickson91a}
R.~P. Erickson and D.~L. Mills, Phys. Rev. B {\bf 43}, 10715 (1991)

\bibitem{Erickson91}
R.~P. Erickson and D.~L. Mills, Phys. Rev. B {\bf 44}, 11825 (1991)

\bibitem{Pereira99}
J.~J.~M.~Pereira and M.~G. Cottam, J. Appl. Phys. {\bf 85}, 4949 (1999)

\bibitem{Rezende09}
S.~M. Rezende, Phys. Rev. B {\bf 79}, 174411 (2009)

\bibitem{Costa98}
R.~N. Costa~Filho, M.~G. Cottam, and G.~A. Farias, Solid State Commun. {\bf
  108}, 439 (1998)

\bibitem{Costa00}
R.~N. Costa~Filho, M.~G. Cottam, and G.~A. Farias, Phys. Rev. B {\bf 62}, 6545
  (2000)

\bibitem{Benson69}
H.~Benson and D.~L. Mills, Phys. Rev. {\bf 178}, 839 (1969)

\bibitem{Ziman79}
J.~M. Ziman, {\em Principles of the Theory of Solids\/}, Cambridge University
  Press, Cambridge (1979)

\bibitem{Kostylev07}
M.~P. Kostylev, G.~Gubbiotti, J.-G. Hu, G.~Carlotti, T.~Ono, and R.~L. Stamps,
  Phys. Rev. B {\bf 76}, 054422 (2007)

\bibitem{Cherepanov93}
V.~Cherepanov, I.~Kolokolov, and V.~L'vov, Physics Reports {\bf 229}, 81 (1993)

\bibitem{Gilleo58}
M.~A. Gilleo and S.~Geller, Phys. Rev. {\bf 110}, 73 (1958)

\bibitem{Tittmann73}
B.~R. Tittmann, Solid State Commun. {\bf 13}, 463 (1973)

\bibitem{Gurevich}
A.~G. Gurevich and G.~A. Melkov, {\em Magnetization Oscillations and Waves\/},
  CRC Press, Boca Raton (1996)

\bibitem{Tupitsyn08}
I.~S. Tupitsyn, P.~C.~E. Stamp, and A.~L. Burin, Phys. Rev. Lett. {\bf 100},
  257202 (2008)

\bibitem{Bartosch06}
L.~Bartosch, L.~Balents, and S.~Sachdev, Ann. Phys. (New York) {\bf 321}, 1528
  (2006)

\end{thebibliography}
\end{document}